\documentclass[12pt]{article}
\usepackage[cp1251]{inputenc}
\usepackage[english]{babel}
\usepackage{amssymb,amsmath}
\usepackage{graphicx}

\title{ Few-Nucleon Systems: Notes about the Status and Results of
Investigations }

\author{Yu.P.Lyakhno}

\begin {document}

\maketitle

\begin{center} {\it National Science Center "Kharkov Institute of
Physics and Technology" \\ 61108, Kharkiv, Ukraine}
\end{center}

\begin{abstract}

The model-independent calculation of the nuclei ground state and the
states of scattering  can be carried out with due regard for
realistic {\it NN} and 3{\it N} forces between nucleons and also,
with the use of exact methods of solving the many-nucleon problem.
The tensor part of {\it NN} interaction and 3{\it NF}'s generate the
lightest nuclei states with nonzero orbital momenta of nucleons.
These states in the lightest nuclei are the manifestation of the
properties of inter-nucleonic forces, and therefore, similar effects
should be observed in all nuclei and also in all their excited
states. In this paper primary attention is given to the
investigation of the $^4$He nucleus.

\vskip20pt

 PACS numbers: 21.30.-x; 21.45.+v; 25.20.-x; 27.90.+b.

\end{abstract}

\section{ Introduction }

From the physical standpoint, to describe the nucleon system, one
must know the nucleon properties and inter-nucleonic forces. The
world constants and nucleon properties are known within sufficient
accuracy, while inter-nucleonic forces are complicated in
character and are known to a less accuracy. Unlike the atom, these
forces cannot be described by the 1/$r^2$  ratio (where r is the
distance between nucleons) or by more complicated expressions like
the Woods-Saxon potential \cite{1}. The distinctive feature of
inter-nucleonic forces is that they depend not only on the
distance {\it r}, but also on the quantum configuration of the
nucleon system, which is determined by the orbital momentum {\it
L}, spin {\it S} and isospin {\it T} of this system.

The {\it NN} potential can be determined phenomenologically from
the experimental data on the ground state of the two-nucleon
system and on the elastic ({\it p,p}), ({\it n,p}) and ({\it n,n})
scattering at nucleon energies up to 500 MeV. At higher nucleon
energies, nonelastic processes come into play, and the potential
approach becomes inapplicable. However, the data about the
inter-nucleonic forces in this nucleon energy region are
sufficient for the description of nucleus ground state, and also
of nuclear reactions up to the meson-producing threshold.
Nowadays,  Argonne AV18 \cite{2} and  CD-Bonn \cite{3} appear to
be the most accurate potentials. In the construction of the
charge-dependent CD-Bonn potential in the range of
laboratory-system nucleon energies up to 350 MeV, 2932 ({\it
p,p})- and 3058 ({\it n,p})- scattering data were used. Adjustable
expressions were derived on the basis of the meson model of strong
interaction of nucleons. Into the account were taken the $\pi$,
$\eta$, $\rho$, $\omega$ -one-meson-exchange contribution,
$\sigma$- one-boson-exchange, 2$\pi$ -exchange, including
$\Delta$-isobar configurations, and of the $\pi$$\rho$-exchange
contribution. The calculations of the {\it NN} potential waves
were up to $J\leqslant$4 (the following notation is used for the
purpose: $^{2S+1}L_J$, {\it J} is the total momentum of the
system). The quantity  $\chi^2$/datum was found to be 1.02.

Mathematically, to describe the nucleon system, it is necessary to
use the accurate methods of solving the many-nucleon problem. To
describe the three-body system in the case of an arbitrary two-body
potential, Faddeev \cite{4} suggested solving a set of connected
integral equations.  Later on, Yakubovsky \cite{5} generalized this
result for the case of any number of particles in the system. At
present, it is found that 3{\it N} forces take action in the
nucleus. In the calculations, the 3{\it N} potentials of types
UrbanaIX \cite{6} and  Tucson-Melbourne \cite{7,8} are most
frequently used. So, for exact description of a many-nucleon system
it is necessary to solve the set of connected integral equations
with due regard for the contribution of {\it NN} and 3{\it N}
forces. The solution of the problem by the Faddeev-Yakubovsky (FY)
method was reported by Gloeckle and Kamada (GK) \cite{9}. The
characteristics of ground states three- and four-nucleon nucleus by
the FYGK method were calculated in papers \cite{10}-\cite{14}.

The realistic {\it NN} and {\it NNN} forces were also used in the
calculations by the Lorentz integral transform (LIT) method
\cite{15}, the hyperspherical harmonic variational method (HHVM)
\cite{16}, the refined resonating group model (RRGM) \cite{17} and
others \cite{18}.

It is hoped that the accuracy of measurements of realistic {\it
NN} and {\it NNN} potentials would get further better, in
particular, at the expense of using the data from
double-polarization experiments \cite{19}. A number of
laboratories create targets of the polarized $^3$He nuclei
\cite{20}-\cite{22}. The investigation of disintegration of
polarized $^3$He nuclei by polarized beams of particles can
provide some new information about  3{\it N} forces.

Along with the elaboration more precise definition of
phenomenological potentials, important results were obtained
through theoretical calculations of inter-nucleonic forces within
the framework of chiral effective field theory (EFT). At present,
the calculation of chiral interactions is not as accurate as that
of phenomenological {\it NN} forces. Calculated within the
framework of the EFT, the {\it NN} potential parameters for
partial waves with J$\leqslant$2 \cite{23} are in satisfactory
agreement with the experiment in the region of nucleon lab energy
up to $T_N\sim$290 MeV . In the context of the EFT, Rozpedzik {\it
et al.} \cite{24} estimated the effect of 4{\it N} forces and
found the additional contribution of 4{\it N} forces to the
binding energy of the $^4$He nucleus to be about several hundreds
of keV. The calculations in the context of EFT are of particular
importance for explaining the origin and explicit representation
of 3{\it N} and 4{\it N} forces. The reason is that there are a
good many experimental data to determine the {\it NN} potential,
whereas for determination of 3{\it N} and 4{\it N} forces these
data are not nearly enough. The origin and the explicit form of
3{\it N} and 4{\it N} forces is a basic issue of few-nucleon
systems.

Section 2 presents the results of theoretical calculations of the
ground states of few-nucleon nuclei. Section 3 presents the
multipole analysis of the $^4$He($\gamma,p$)T and
$^4$He($\gamma,n$)$^3$He reactions, performed on the basis of the
experimental data about the differential cross-sections and
cross-section asymmetry with linearly polarized photons. The
possible effects, determined by realistic inter-nucleonic forces
in nuclei with $A >$ 4 are discussed in Section 4. The conclusions
are formulated in Section 5.

\section{Results of the theoretical calculations}

In Ref.\cite{11} calculation of the ground-state of the
$\alpha$-particle is carried out. The calculations took into
account the contributions from the states of the {\it NN} system
having the total momentum up to $J\le$6. The consideration of
large total-momentum values of the two-nucleon system is
necessary, for example, for a correct calculation of short-range
correlations. In the calculation \cite{11}, account was taken of
the states, in which the algebraic sum of orbital momenta of all
nucleons of the $^4$He nucleus was no more than $l_{max}$=14. The
system comprised 6200 partial waves. The authors of work \cite{11}
estimated their mistake in the calculations of $^4$He nuclear
binding energy to be $\sim$50 keV. Considering that the calculated
value of the binding energy is $\sim$200 keV higher than the
experimental value, the authors have made a conclusion about a
possible contribution of 4{\it N} forces that are of repulsive
nature.

Table 1 lists the values of nuclear binding energies (in MeV) for
$^4$He, $^3$H, $^3$He and $^2$H, calculated with the use of {\it
NN} potentials  AV18 and 3{\it NF}'s  UrbanaIX.  It is evident
from the table that without taking into account the 3{\it N}
forces, the nuclei appear underbound, while with due regard for
the forces the agreement with experimental data is satisfactory.

{\bf T a b l e 1: Binding energies (in MeV units) of $^4$He, of
$^3$H, of $^3$He and of $^2$H, calculated with  Argonne V18 and
Argonne V18 + Urbana IX interaction.}
\begin{center}
\begin{tabular}[t]{|c|c|c|c|c|c|}
\hline {Inter-} & & & & & \\
 {action}& Method & $^4$He & $^3$H
& $^3$He & $^2$H \\
\hline
 AV18 & FY & -24.28 & -7.628 & -6.924  & \\
&  RRGM & -24.117 & -7.572 & -6.857 & -2.214 \\ & HHVM & -24.25 &
&  &   \\ \hline  AV18+ & FYGK & -28.50  & -8.48 & -7.76 & \\
+UIX& RRGM & -28.342 & -8.46 & -7.713 & -2.214   \\ & HHVM &
-28.50 & -8.485 & -7.742 &  \\ \hline &  Exp & -28.296 & -8.481 &
-7.718 & -2.224 \\ \hline
\end{tabular}
\end{center}

Similar results were obtained with the use of the {\it NN}
potential CD-Bonn and the 3{\it N} potential  Tucson-Melbourne.

Table 2 gives the calculated root-mean-square radii $r_{rms}$ of
the $^4$He nucleus \cite{16,17}. The agreement with experiment is
also satisfactory.

{\bf T a b l e 2: The $^4$He nucleus
$<r^2>^{1/2}$ radii (fm)}.
\begin{center}
\begin{tabular}[t]{|c|c|c|}
\hline
$\quad$Interaction$\quad$ & $\quad$Method$\quad$ &$\quad$ $^4$He$\quad$  \\
\hline
 AV18 & RRGM & 1.52  \\
 & HHVM & 1.512  \\
\hline
 AV18+UIX & RRGM & 1.44 \\
& HHVM & 1.43  \\ \hline & Exp & 1.67 \\ \hline
\end {tabular}
\end{center}

It should be also noted that the Coulomb interaction between protons
results in the production of {\it T}=1 and {\it T}=2 isospin states
of $^4$He. Table 3 gives the probabilities of these states for the
$^4$He nucleus calculated in papers \cite{11}, \cite{16}.

\vskip 50pt

{\bf T a b l e 3: Contribution of different total isospin states to
the $^4$He nuclear wave function. The values are given in \%.}
\begin{center}
{\begin{tabular}[t]{|c|c|c|c|c|} \hline
Interaction & Method & T=0 & T=1 & T=2  \\
\hline
AV18 & FY & 99.992 & 3·10$^{-3}$ & 5·10$^{-3}$ \\
& HHVM & & 2.8·10$^{-3}$ & 5.2·10$^{-3}$ \\
\hline
\end {tabular}}
\end{center}
The tensor part of {\it NN} interaction and 3{\it NF's} generate
the $^4$He nuclear states with nonzero orbital momenta of
nucleons. Table 4 gives the probabilities of {\it S},
$S^{\prime}$, {\it P} and {\it D} states of the $^4$He and $^3$He
nuclei calculated by Nogga {\it et al.} \cite{11}, where
$S^{\prime}$ is a part of $^1S_0$-states with nonzero orbital
momenta of nucleons. The calculations gave the probability of
$^5D_0$ states having the total spin {\it S}=2 and the total
nucleon orbital momentum {\it L}=2 of the $^4$He nucleus to be
$\sim16\%$, and the probability of $^3P_0$ states having {\it
S}=1, {\it L}=1 to be between 0.6\% and 0.8\%. It is obvious from
Table 4 that the consideration of the 3{\it NF's} contribution
increases the probability of $^3P_0$ states by a factor of
$\sim$2.

{\bf T a b l e 4: S, $S^{\prime}$, P, and D state
probabilities for $^4$He and $^3$He.}
\begin{center}
\begin{tabular}[t]{|c|c|c|c|c|c|c|c|c|} \hline
& \multicolumn{4}{|c|}{$^4$He} & \multicolumn{4}{|c|}{$^3$He}  \\
\cline{2-9}
Interaction &  $S\%$ &  $S^{\prime}\%$ & $ P\%$ & $ D\%$  & $S\%$ &  $S^{\prime}\%$ & $ P\%$ & $ D\%$ \\
\hline
 AV18 & 85.45 & 0.44 & 0.36 & 13.74 & 89.95 & 1.52 & 0.06 & 8.46 \\
\hline
 CD-Bonn & 88.54 & 0.50 & 0.23 & 10.73 & 91.45 & 1.53 & 0.05 & 6.98 \\
\hline
 AV18+UIX & 82.93 & 0.28 & 0.75 & 16.04 & 89.39 & 1.23 & 0.13 & 9.25 \\
\hline
 CD-Bonn+TM & 89.23 & 0.43 & 0.45 & 9.89 & 91.57 & 1.40 & 0.10 & 6.93 \\
\hline
\end{tabular}
\end{center}

{\section{The multipole analysis of $^4$He($\gamma,p$)$^3H$ and
$^4$He($\gamma,n$)$^3$He reactions}}

In the {\it E1}, {\it E2} and {\it M1} approximations, the laws of
conservation of the total momentum and parity for two-body
$(\gamma,p)$ and $(\gamma,n)$ reactions of $^4$He nuclear
disintegration permit two multipole transitions $E1^1P_1$ and
$E2^1D_2$ with the spin {\it S}=0 and four transitions $E1^3P_1$,
$E2^3D_2$, $M1^3S_1$ and $M1^3D_1$ with the spin {\it S}=1 of
final-state particles. The differential cross section in the c.m.s.
can be expressed in terms of multipole amplitudes as follows
\cite{25,26}:

\newcommand{\lambcross}{\lambda \hspace{-0.6em} {^-}}
{ \begin{eqnarray}\label{eq1} &&\frac{\rm d\sigma }{\rm d\Omega}=
\frac{ \lambcross^2}{32}\{\sin^2\theta[18|E1^1{\rm P}_1|^2-
9|E1^3{\rm
P}_1|^2\nonumber\\
&&+9|M1^3{\rm D}_1|^2-25|E2^3{\rm D}_2|^2\nonumber\\
&&-18\sqrt2Re(M1^3{\rm S}_1^*\,M1^3{\rm D}_1)+ 30\sqrt3Re(M1^3{\rm
D}_1^*\,E2^3{\rm D}_2)\nonumber\\
&&+30\sqrt6Re(M1^3{\rm S}_1^*\,E2^3{\rm D}_2)\nonumber\\
&&+\cos\theta(60\sqrt3Re(E1^1{\rm P}_1^*\,E2^1{\rm D}_2)\nonumber\\
&&-60 Re(E1^3{\rm P}_1^*\,E2^3{\rm D}_2))\nonumber\\
&&+\cos^2\theta(150| E2^1{\rm D}_2|^2-100| E2^3{\rm D}_2|^2)]\nonumber\\
&&+\cos\theta[-12\sqrt6Re(E1^3{\rm P}_1^*\,M1^3{\rm S}_1)\nonumber\\
&&-12\sqrt3Re(E1^3{\rm P}_1^*\,M1^3{\rm D}_1)+60Re(E1^3{\rm
P}_1^*\,E2^3{\rm D}_2]\nonumber\\
&&+18| E1^3{\rm P}_1|^2+12| M1^3{\rm S}_1|^2+6| M1^3{\rm D}_1|^2\nonumber\\
&&+50| E2^3{\rm D}_2|^2+12\sqrt2Re(M1^3{\rm S}_1^*\,M1^3{\rm D}_1)\nonumber\\
&&-20\sqrt6Re(M1^3{\rm S}_1^*E2^3{\rm D}_2)-20\sqrt3Re(M1^3{\rm
D}_1^*\,E2^3{\rm D}_2)\}\,,\nonumber\\
\end{eqnarray}}
where $\lambcross$ is the reduced wavelength of the photon.

It is known that the cross-section asymmetry of the linearly
polarized photon reaction is described by the following expression
\cite{27}: {\begin{eqnarray}\label{eq2} &&{\Sigma }({\theta })=
\sin^2\theta\{18 \mid E1^1{\rm P}_1|^2-9|
E1^3{\rm P}_1|^2\nonumber\\
&&-9| M1^3{\rm D}_1|^2+25| E2^3{\rm D}_2|^2\nonumber\\
&&+18\sqrt2Re(M1^3{\rm S}_1^*\,M1^3{\rm D}_1)+ 10\sqrt3Re(M1^3{\rm
D}_1^*\,E2^3{\rm D}_2)\nonumber\\
&&+10\sqrt6Re(M1^3{\rm S}_1^*\,E2^3{\rm D}_2)\nonumber\\
&&+\cos\theta[60\sqrt3Re(E1^1{\rm P}_1^*\,E2^1{\rm D}_2)\nonumber\\
&&-60Re(E1^3{\rm P}_1^*\,E2^3{\rm D}_2)]\nonumber\\
&&+\cos^2\theta[150|E2^1{\rm D}_2|^2-100|E2^3{\rm
D}_2|^2]\}/\frac{32}{\lambcross^2}\frac{\rm d\sigma }{\rm d\Omega
}\,.
\end{eqnarray}}
The differential cross section can be presented as:
{\begin{equation} \label {eq3} \frac{\rm d\sigma }{\rm d\Omega }=
A[{\sin^2\theta(1+\beta\cos\theta
+\gamma\cos^2\theta)+\varepsilon\cos\theta+\nu}]\,.\quad
\end{equation}}
In the same terms, the cross-section asymmetry of the linearly
polarized photon reaction can be represented as follows:
{\begin{equation} \label {eq4}
 {\Sigma }({\theta }) = \frac
 {\sin^2\theta(1+\alpha+\beta\cos\theta+\gamma\cos^2\theta)}
 {\sin^2\theta(1+\beta\cos\theta+\gamma\cos^2\theta)+\varepsilon\cos\theta+\nu}.
\end{equation}}
The coefficients A, $\alpha$, $\beta$, $\gamma$, $\varepsilon$, and
$\nu$ are unambiguously connected with multipole amplitudes. As it
is obvious from relation (3), only 5 independent coefficients can be
calculated in the long-wave approximation using the data on the
differential reaction cross-section. So, an improvement in the
accuracy of measuring only the differential reaction cross-section
gives no way of obtaining information about subsequent multipole
amplitudes. In this case, the number of unknown parameters in the
right side of eq. (1) would increase much quicker than the number of
found coefficients in the left side of the equation. In this
connection, in order to obtain information on the succeeding
multipole amplitudes, polarization experiments or other data sources
are required. As it can be seen from relation (4), the experimental
data on the asymmetry of the linearly polarized photon reaction
cross-section enable one to calculate the sixth independent
coefficient.

It can be demonstrated that on the assumption that
$\sigma$($E2^3{\rm D}_2)\gg \sigma(M1)$, from expressions (1) and
(2) we obtain: {\begin {equation}\label{eq5}
 \alpha = \frac
{50 | E2^3{\rm D}_2 | ^2}{18| E1^1{\rm P}_1 | ^2-9| E1^3{\rm P}_1 |
^2-25| E2^3{\rm D}_2 | ^2}>0\,.\quad
\end {equation}}
If we assume that $\sigma(M1)$$\gg \sigma$($E2^3{\rm D}_2$), then we
have {\begin{eqnarray}\label{eq6} \alpha =\left\{-18 | M1^3{\rm
D}_1|^2 \right.\qquad\qquad\qquad\qquad\qquad\qquad\nonumber\\
+36\sqrt2 | M1^3{\rm S}_1 || M1^3{\rm D}_1 | \cos[\delta(^3{\rm
S}_1) - \delta(^3{\rm D}_1)]\}/ \nonumber\\
\{18 | E1^1{\rm P}_1|^2-9 | E1^3{\rm P}_1|^2+9 | M1^3{\rm D}_1|^2 \qquad\nonumber\\
-18\sqrt2| M1^3{\rm S}_1 || M1^3{\rm D}_1 | \cos[\delta(^3{\rm S}_1)
- \delta(^3{\rm D}_1)]\}\,.\quad
\end{eqnarray}}
From the phase analysis of elastic (p,$^3$He) scattering Murdoch
{\it et al.} \cite{28} have determined the phase difference to be
$\delta(^3{\rm S}_1) - \delta(^3{\rm D}_1)>90^0$. Therefore, the
both components in the numerator of expression (6) enter with the
minus sign, and the coefficient $\alpha$ must be negative.

The angular dependence of cross-section asymmetry in the
$^4$He($\vec\gamma,p$)T and $^4$He($\vec\gamma,n$)$^3$He reactions
with linearly polarized photons of energies 40, 56 and 78 MeV was
measured by Lyakhno {\it et al.} \cite{27,29}. The beam of
linearly polarized photons was produced as a result of coherent
bremsstrahlung of 500, 600 and 800 MeV electrons, respectively, in
a thin diamond single crystal. The reaction products were
registered with the use of a streamer chamber located in the
magnetic field \cite{30}. The observed data on the angular
dependence of the cross-section asymmetry are presented in Fig.1.
Here the square represents the data obtained with semi-conductor
detectors by the $\Delta$E-E method \cite{31}. The dashed curve is
the calculation by Mel'nik and Shebeko \cite{32} made in the
plane-wave impulse approximation with consideration of the direct
reaction mechanism and the mechanism of recoil. The solid curves
represent the calculation \cite{33} that meets the requirements of
covariance and gauge invariance. The calculation took into account
the contribution of a number of diagrams corresponding to the pole
mechanisms in s-, t- and u-channels, the contact diagram c, and
also a number of triangular diagrams. A satisfactory fit of the
calculation to the experiment confirms an essential role of the
direct reaction mechanism, the mechanism of recoil and the
final-state rescattering effects.

\begin{figure}[h]
\noindent\centering{
\includegraphics[width=80mm]{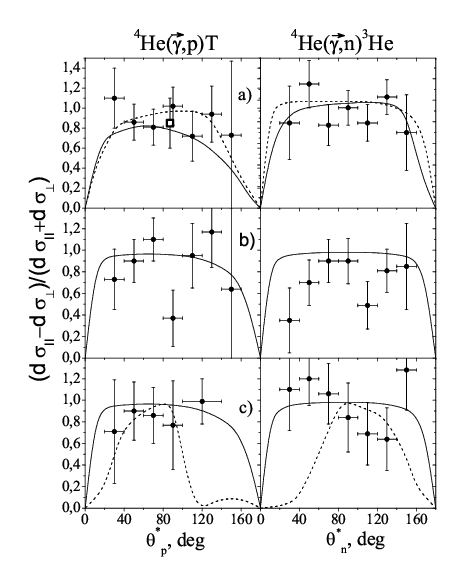}
} \caption{Angular dependence of cross-section asymmetry
$^4$He($\vec\gamma,p$)$^3$H and $^4$He($\vec\gamma,n$)$^3$He
reactions with linearly polarized photon. The points represent the
results of Refs. [27, 29]: a) $E_{\gamma}^{peak}$=40 MeV, b)
$E_{\gamma}^{peak}$=56 MeV, c) $E_{\gamma}^{peak}$=78 MeV. The
square shows the data of Ref. \cite{31}. The errors are
statistical. The solid curve - from Ref. \cite{33}, the dashed
curve - from Ref. \cite{32}.}
\end{figure}

\begin{figure}[h]
\noindent\centering{
\includegraphics[width=80mm]{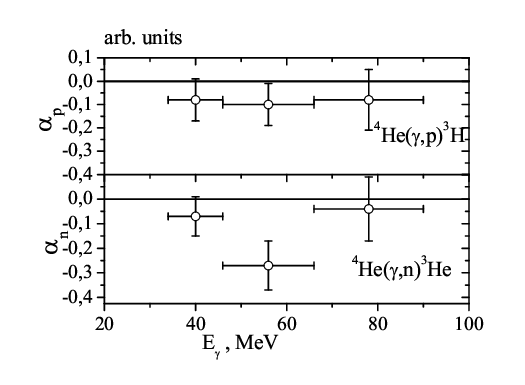}
} \caption{Coefficients $\alpha_p$ and $\alpha_n$. The errors are
statistical only.}
\end{figure}

As a result of the least-squares fit of expressions (3) and (4) to
the experimental data on the differential cross section
\cite{34,35} and cross-section asymmetry of linearly polarized
photon reactions, the coefficients A, $\alpha$, $\beta$, $\gamma$,
$\varepsilon$, and $\nu$ were calculated \cite{27}. Since the
coefficients enter into relations (3) and (4) in linear fashion,
the solution was unambiguous.

Since only phase differences enter into formulas (1) and (2), these
relations comprise 11 unknown parameters. The currently available
experimental data on the ($\gamma,p$) and ($\gamma,n$) reactions are
insufficient for determining all the parameters. According to the
experimental data obtained (see Fig.2), $\alpha_p$ and $\alpha_n$
are the minus coefficients, and hence, the least amplitude that
enters into expressions (1) and (2) is the E2$^3D_2$ amplitude.
After the E2$^3D_2$-comprising components are excluded, expressions
(1) and (2) still comprise 9 unknown parameters: $|E1^1P_1|$,
$|E2^1D_2|$, $\cos[\delta(^1P_1)-\delta(^1D_2)]$, $|E1^3P_1|$,
$|M1^3S_1|$, $|M1^3D_1|$, $\cos[\delta(^3S_1)-\delta(^3D_1)]$,
$\cos[\delta(^3S_1)-\delta(^3P_1)]$ and
$\cos[\delta(^3P_1)-\delta(^3D_1)]$.

It is known \cite{36} that according to the isospin selection
rules for self-conjugate nuclei the isoscalar parts of {\it E1}
and {\it M1} amplitudes are essentially suppressed. In view of
this, using the Watson theorem, the last three phase differences
were calculated from the data of phase analysis of elastic
(p,$^3$He) scattering \cite{28}.

The coefficients $A$, $\alpha$, $\beta$, $\gamma$, $\varepsilon$,
and $\nu$ are expressed in terms of the multipole amplitudes as: {
\begin{eqnarray}\label{eq7} A = \lambcross^2/32\{18 | E1^1{\rm
P}_1|^2-9| E1^3{\rm P}_1|^2+9|
M1^3{\rm D}_1|^2 \nonumber\\
-18\sqrt2|M1^3{\rm S}_1||M1^3{\rm
D}_1|\cos[\delta(^3S_1)-\delta(^3D_1)]\};
\end{eqnarray}}
{\begin{eqnarray}\label{eq8} \alpha =\{-18 | M1^3{\rm
D}_1|^2 \qquad\qquad\qquad\qquad\qquad\qquad\nonumber\\
+36\sqrt2 | M1^3{\rm S}_1 || M1^3{\rm D}_1 | \cos[\delta(^3{\rm
S}_1) - \delta(^3{\rm D}_1)]\}\,/\,\frac{32}{\lambcross^2}A\,;
\end{eqnarray}}
{\begin{equation}\label{eq9} \beta =60\sqrt3|E1^1{\rm P}_1||E2^1{\rm
D}_2|\,cos[\delta(^1P_1)-\delta(^1D_2)]\,/\,\frac{32}{\lambcross^2}A\,;
\end {equation}}
{\begin {equation}\label{eq10} \gamma=150|E2^1{\rm
D}_2|^2\,/\,\frac{32}{\lambcross^2}A\,;\qquad\qquad\qquad\qquad\qquad
\end {equation}}
{\begin{eqnarray}\label{eq11} \varepsilon =\{-12\sqrt3|E1^3{\rm
P}_1||M1^3{\rm
D}_1|\,cos[\delta(^3P_1)-\delta(^3D_1)] \nonumber\\
-12\sqrt6|E1^3{\rm P}_1||M1^3{\rm
S}_1|\,cos[\delta(^3P_1)-\delta(^3S_1)]\}\,/\,\frac{32}{\lambcross^2}A\,;\nonumber\\
\end{eqnarray}}
{\begin{eqnarray}\label{eq12} \nu =\{18 | E1^3{\rm P}_1|^2+12|
M1^3{\rm S}_1|^2 +6| M1^3{\rm
D}_1|^2 \qquad\nonumber\\
+12\sqrt{2}|M1^3{\rm S}_1||M1^3{\rm
D}_1|\,cos[\delta(^3S_1)-\delta(^3D_1)]\}\,/\,\frac{32}{\lambcross^2}A.\nonumber\\
\end{eqnarray}}
The experimentally observable quantities are expressed in terms of
multipole amplitudes in a bilinear fashion. Therefore, there must
exist two different sets of multipole amplitudes, which satisfy
these experimental data. With the help of programs of the least
square method (LSM) one positive solution of the problem can be
calculated. The second solution can be found, for example, by the
lattice method. Since both positive solutions have the equal
$\chi^2$ values, an additional information is necessary to choose
the proper solution. It should be also noted that if the difference
between the solutions is comparable with the amplitude errors, then
the LSM errors of the amplitudes may appear overestimated.

The amplitude values were calculated from the derived set of six
bilinear equations (7-12) with six unknown parameters using the
random-test method \cite{27}. To calculate the errors in the
amplitudes, 5000 statistical samplings of A, $\alpha$, $\beta$,
$\gamma$, $\varepsilon$, and $\nu$ values with their errors were
performed. The errors in the coefficients were assumed to be
distributed by the normal law. After each statistical sampling the
set of equations was solved, the calculated amplitude values were
stored and then their average values and dispersions were
calculated.

According to Ref. \cite{37}, one can assume that with the photon
energy increase the   $M1^3S_1$ transition cross-section decreases
as 1/{\it V},  where {\it V} is the nucleon velocity. Therefore,
at MeV nucleon energies the contribution of the M1$^3S_1$
transition can be neglected. In this connection, out of the two
found solutions of the system of equations (7-12) the choice has
been made on the solution, where $\sigma(E1^3P_1)>
\sigma(M1^3S_1)$.

\begin{figure}[h]
\noindent\centering{
\includegraphics[width=80mm]{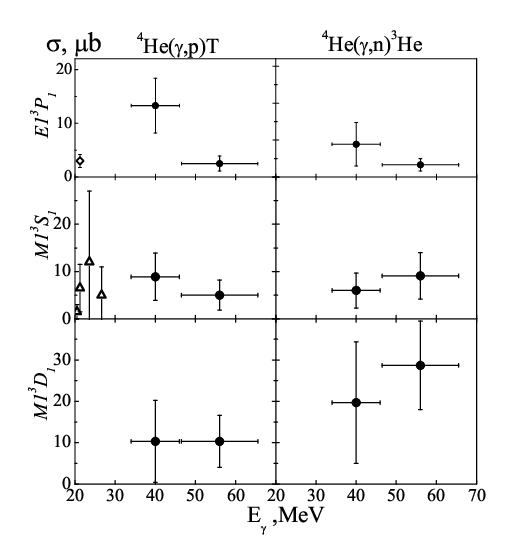}
 \caption{Total cross sections of spin {\it S}=1 transitions
of $^4$He($\gamma,p$)$^3$H and $^4$He($\gamma,n$)$^3$He reactions:
$\Diamond$-data from Ref. \cite{37}; $\triangle$-data of Ref.
\cite{38}; $\bullet$-data of Ref. \cite{27}. The errors are
statistical only.}}
\end{figure}

The findings of the experiment aimed to determine the total cross
sections of {\it S}=1 transitions are presented in Fig.3. The
triangles represent the data of Wagenaar {\it et al.} \cite{38},
the diamond shows the data of Pitts \cite{37} obtained from
studies of the reaction of radiative capture of protons by tritium
nuclei. The points represent the data of Lyakhno {\it et al.}
\cite{27} from the studies of two-body ($\gamma,p$) and
($\gamma,n$) reactions of $^4$He disintegration. The existing
experimental data on the total cross-sections of electromagnetic
transitions with the spin {\it S} =1 in the $^4$He($\gamma,p$)T
and $^4$He($\gamma,n$)$^3$He reactions have considerable statistic
and systematic errors.

{\section{Role of the spin-orbit interaction in nuclei}}

The occurrence of states with nonzero orbital momenta of nucleons
in the lightest nuclei is a manifestation of the properties of
inter-nucleonic forces and, hence, such effects should be observed
without exception in all nuclei as well as in all their excited
states.

One can suppose that the contribution of the effects connected
with the tensor part of {\it NN} potential and 3{\it NF's}
increases with a growth of the atomic number. Firstly, it can be
seen from the fact that the {\it D}-state contribution in the
deuteron is about 5\%, while in the $^4$He nucleus it makes
$\sim$16\%.

Secondly, at the calculation of the probabilities of the outside
shell states, for example, in the $^{12}$C nucleus, similarly to
the case with the $^4$He nucleus, we must bear in mind that the
nucleus spin of $^{12}$C can take the values 0$\le S \le $6, the
total orbital momentum of the nucleons of $^{12}$C can be 0$\le L
\le $6, and the orbital momenta of separate nucleons can take on
any values, which are not forbidden by the Pauli principle. In
other words, the ground state of the nucleus $^{12}$C can be of
the $^1S_0$, $^3P_0$, $^5D_0$, $^7F_0$, $^9G_0$, $^{11}H_0$, or
$^{13}I_0$ states. Nowadays, probability of these states is not
held due to their extreme complication. However, the rough
estimation of these probabilities can be achieved in the following
way. Let us suppose the $^{12}$C nucleus consist of three weakly
bound $\alpha$-clusters. The total momentum and parity
conservation laws do not forbid, and the tensor part of {\it NN}
interaction and 3{\it NF's} initiate states with the orbital
momenta larger than predicted by the nuclear shell model (NSM)
\cite{39,40} in every cluster, in two clusters at one time or in
all three clusters. In the result, the probability of these states
in the $^{12}$C can be higher than in the $^4$He nucleus.

Using this supposition one can explain the row of the well-known
nuclei properties. Particularly, it is possible to explain the
significantly higher contribution of the spin-orbit interaction in
the nuclei, than calculated in the NSM frames. Contribution of the
spin-orbit interaction {\it A} nucleons in the potential energy
nucleus can be assessed by the relation:
\begin{equation}
\label {eq13}
 U_{SO}=-\lambda\bigg(\frac{\hbar}{Mc}\bigg)^2\sum_{i=1}^A\frac{1}{r_i}\frac{\partial V_i}{\partial r_i}
 (\vec l_i \cdot\vec s_i)\,,
\end{equation}
where $M$ is the nucleon mass, $V_i$ is the spherically
symmetrical potential, {\it l} is the orbital momentum,  {\it s}
is the spin of the nucleon. However, for the agreement of the
experimental data into the expression (13) was put a constant,
which is $\lambda\sim$10.

The appearance of the fitting constant $\lambda$  can be partially
explained as follows. Let {\it k}-number of the nucleons,  with the
orbital momenta in accordance with the NSM, and other {\it A-k}
nucleons have orbital momenta bigger than it is predicted by the
NSM. Then the expression (13) can be refined as:
\begin{eqnarray}\label {eq14}
U_{SO}=-\left(\frac{\hbar}{Mc}\right)^{2}\left[\sum_{i=1}^k
\frac{1}{r_i}\frac{\partial V_i}{\partial r_i}
 P(l_i^{sh})(\vec l_i^{sh} \cdot\vec s_i)\right.\nonumber\\
 \left.+\sum_{i=k+1}^A \frac{1}{r_i}\frac{\partial
V_i}{\partial r_i}
 P(l_i>l_i^{sh})(\vec l_i \cdot\vec s_i)\right]\,,
\end{eqnarray}
where $P(l_i)$ is the probability for the {\it i}-th nucleon to
have the orbital momentum $l_i$. The sum of probabilities is
$\sum_{i=1}^A P({l_i})$=1. Thus, in the case of the lightest
nuclei second summing of the expression (14) leads to the small
but not equal zero impact of spin-orbit interaction to the
potential energy of the nucleus. In medium and heavy nuclei
nucleons are situated, generally, in $l>l^{sh}$ states. In other
words, the tensor part of {\it NN} potential and 3{\it N} forces
push the nucleons outside of nuclear shells, with the rise of
atomic number the role of this effects is rising at that. The
second summing, which is not predicted by NSM, can give a
significant additional contribution to potential energy of a
nucleus.

In particular, calculations \cite{41,42}, made on the basis of the
Woods-Saxon potential, can give the overestimation of the protons
number {\it Z} for the position of the island of stability  of the
superheavy nuclei. In the frames of semiempirical shell models
\cite{43} at the extrapolation of fitting expressions for the area
of the heavy nuclei to the area of the superheavy nuclei,
apparently, the according corrections should be also counted

\section{Conclusions}

Modern methods of the decision of a many-nucleon problem make it
possible to calculate the characteristics of this nucleus to an
accuracy, which is determined by the accuracy the measurement of
{\it NN} potential, and 3{\it NF,s} forces. In this connection the
$^4$He nucleus is an ideal laboratory for investigating the
properties of these forces. The measurement of total cross-sections
for the electromagnetic transitions with spin {\it S}=1 in the final
state of the particle system in $^4$He($\gamma,p$)$^3$H and
$^4$He($\gamma,n$)$^3$He reactions, and, in addition to data about
radiative deuteron-deuteron capture, can  give new information about
of the structure of $^4$He nucleus.

The states with non-zero orbital momenta of the nucleons of the
lightest nuclei are the manifestation of the properties of
inter-nucleonic forces and, consequently, such effects should be
observed in all nuclei and in all their excited states without any
exception. One can propose, that the tensor part of {\it NN}
potential and 3{\it N} forces push the nucleons outside of nuclear
shells, with the rise of atomic number the role of this effects is
rising at that. This can lead to the additional contribution of
the spin-orbital interaction to the potential energy of the
nucleus. In particular, calculations, made on the basis of the
Woods-Saxon potential, can give the overestimation of the protons
number {\it Z} for the position of the "stability island" of the
superheavy nuclei.

Author gives the gratitude to Yu.P. Stepanovsky for important
advice and discussion over the article material, and to A.V.
Shebeko for the sequence of critical remarks.

\end{document}